\documentclass[12pt]{article}
\usepackage{amsmath}
\usepackage{epsfig,graphicx}

\def\be{\begin{equation}}
\def\ee{\end{equation}}
\def\bea{\begin{eqnarray}}
\def\eea{\end{eqnarray}}
\def\tr{{\rm Tr}\:}

\def\rme{\mathrm{e}}
\def\rmi{\mathrm{i}}
\def\scq{{\textsc q}}

\begin{document}

\hfill CCNY-HEP-15-01
\begin{center}
{\Large \bf Statistics of two-dimensional random walks, \\
the ``cyclic sieving phenomenon'' \\
and the Hofstadter model \\}\vskip 0.5cm

{\large \bf Stefan Mashkevich}\footnote{mash@mashke.org}\\[0.1cm]
Schr\"odinger, 120 West 45th St., New York, NY 10036, USA \\ and \\
Bogolyubov Institute for Theoretical Physics, 03143 Kiev, Ukraine\\[0.4cm]
{\large \bf St\'ephane Ouvry}\footnote{ouvry@lptms.u-psud.fr}\\[0.1cm]
Physics Department, City College of the CUNY, New York, NY 10031, USA \\ and \\
Laboratoire de Physique Th\'eorique et Mod\`eles Statistiques,
CNRS-Universit\'e Paris Sud, Facult\'e des Sciences d'Orsay,
91405 Orsay, France
\\[0.4cm]
{\large \bf Alexios Polychronakos}\footnote{alexios@sci.ccny.cuny.edu}\\[0.1cm]
Physics Department, City College of the CUNY, New York, NY 10031, USA \\and \\
The Graduate Center, CUNY, New York, NY 10016 
\vskip 0.4cm
\today
\end{center}

\vskip 0.5cm
\centerline{\large \bf Abstract}
\vskip 0.2cm
We focus on  the algebraic area probability distribution of planar random walks on a square lattice with $m_1, m_2, l_1$ and $l_2$ steps right, left, up and  down. We aim, in particular,
at the algebraic area  generating function $Z_{m_1,m_2,l_1,l_2}(\scq)$ 
evaluated at  $\scq=\rme^{2\rmi\pi\over q}$, a root of unity, when both $m_1-m_2$ and $l_1-l_2$ are multiples of $q$.
In the simple case of staircase walks, a geometrical interpretation
of $Z_{m,0,l,0}(\rme^\frac{2\rmi\pi}{q})$  in terms of the cyclic sieving phenomenon
is illustrated.
Then, an  expression for $Z_{m_1,m_2,l_1,l_2}(-1)$, which is relevant to the Stembridge's case, is proposed.
Finally, the related problem of evaluating 
the $n$th moments of the Hofstadter Hamiltonian in the commensurate case is addressed. 

\section{Introduction}
The cyclic sieving phenomenon \cite{sieving} is quite ubiquitous   in combinatorics,  whereby
some finite sets have both a cyclic symmetry and a generating function which,
when evaluated at roots of unity, happens to count some symmetry classes of the sets.
A paramount example is the collection of the $l$-subsets
of a $(m+l)$-set with the $\scq$-binomial  generating function
\be\label{start}  
{m+l\choose l}_{\scq}  \equiv \frac{[m+l]_{\scq}!}{[m]_{\scq}! [l]_{\scq}!} \;,
\ee
The {\scq}-factorial is defined as
\be 
[l]_{\scq}!  =  \prod_{i=1}^{l}{1-{\scq}^{i}\over 1-{\scq}}
 =  1 (1+{\scq}) (1+{\scq}+{\scq}^2) \cdots (1 + {\scq} + \ldots + {\scq}^{l-1}) \;.
\ee
Denoting by $c$ the cycling generator
(one-step cyclic permutation of the $(m+l)$ set),
one finds that ${m+l\choose l}_{\scq}$ evaluated at $\scq=e^{2\rmi\pi p/(m+l)}$
counts the number of $l$-subsets fixed by (i.e., invariant with respect to)
$c^p$ for $p=1,2,\ldots,m+l$.
In the particular case $\scq=-1$ one refers to the Stembridge's phenomenon \cite{stembridge}.
The fact that integers show up here stems from a well-known identity:
for any integers $p$,  $q$  mutually prime\footnote{In the sequel, $\scq$  denotes
the argument of the $\scq$-binomial, whereas the integer $q$  denotes
the root of unity $\scq=\exp(2\rmi\pi/q)$ at which the generating functions are evaluated.} and  $m$ a multiple of $q$, one has
\be\label{identity}
\binom{m + l}{m}_{\rme^{\frac{2 \rmi \pi  p}{q}}} =
\binom{\left[\frac{m + l}{q}\right]}{\frac{m}{q}} \;.
\ee

A planar random walk on a square lattice is defined as an ordered sequence
of steps to the right $x$, left $x^{-1}$, up $y$, and down $y^{-1}$, with   numbers of  steps
 $m_1$, $m_2$, $l_1$ and $l_2$, respectively.
We will refer to such a walk as an $(m_1,m_2,l_1,l_2)$ walk.
The walk is closed if $m_1 = m_2=m$ and $l_1 = l_2=n/2-m$,
where $n$ is the total number of steps (which is necessarily even),
and open otherwise.
We extend the  standard definition of the algebraic area 
enclosed by a closed walk onto an open walk, as follows:
close an open walk by connecting its end point with its start point,
adding on to the end of the walk the minimum necessary number of steps,
first vertical, then horizontal. E.g., if $m_1 \ge  m_2$ and $l_1 \ge l_2$, then
we close the walk by adding $l_1 - l_2$ steps down followed by
$m_1 - m_2$ steps to the left. The area of the open walk is defined as the area
of the closed walk thus obtained; essentially, it is the algebraic area
{\em under} the open walk.

The generating function $Z_{m_1, m_2, l_1, l_2}(\scq)$ of the algebraic area probability distribution
of  $(m_1,m_2,l_1,l_2)$ walks  originating from a given point on the lattice is defined in terms of  the number $C_{m_1, m_2, l_1, l_2}(A)$ of such walks
 with algebraic area $A$:
\be
Z_{m_1, m_2, l_1, l_2}(\scq) = \sum_{A=-\infty}^\infty C_{m_1, m_2, l_1, l_2}(A)\, \scq^A \;.
\label{defZ}
\ee
$Z_{m_1, m_2, l_1, l_2}(\scq)$ can be viewed as a generalization
of the $\scq$-binomial coefficient: If $x$ and $y$
(identified with steps to the right and up, respectively) satisfy
the commutation relation $xy = \scq yx$, then 
\be
(x + y + x^{-1} + y^{-1})^n =
\sum_{\stackrel{\scriptstyle m_1,m_2,l_1,l_2}{m_1+m_2+l_1+l_2 = n}}
Z_{m_1,m_2,l_1,l_2}(\scq) y^{-l_1}y^{l_2} x^{m_1} x^{-m_2}  \;.
\label{qbingen}
\ee
It can be seen easily that $Z_{m_1, m_2, l_1, l_2}(\scq)$ satisfies the recurrence relation
\bea
Z_{m_1,m_2,l_1,l_2}(\scq) & = &  Z_{m_1,m_2,l_1-1,l_2}(\scq) + Z_{m_1,m_2,l_1,l_2-1}(\scq) \nonumber \\
&& {} + \scq^{l_2-l_1}Z_{m_1-1,m_2,l_1,l_2}(\scq) + \scq^{l_1-l_2}Z_{m_1,m_2-1,l_1,l_2}(\scq) \;,
\label{recbis}
\eea
with the initial condition $Z_{0,0,0,0}({\scq}) = 1$.
It is therefore a polynomial in $\scq$ and $\scq^{-1}$,
the solution of Eq.~(\ref{recbis}), but whose explicit expression remains unknown.

Trivially, when evaluated at $\scq=1$,
\be
Z_{m_1,m_2,l_1,l_2}(1) = {m_1 + m_2 + l_1 + l_2 \choose m_1,m_2,l_1,l_2}
\ee
is the usual multinomial coefficient that counts the  number of $(m_1,m_2,l_1,l_2)$ walks. 

Less trivially, for ``biased walks'', i.e.,
ones that can go right, up and down, but not left ---
that is to say with $m_2=0$ --- the generating function $Z_{m_1,0,l_1,l_2}(\scq)$
has been found \cite{MO09} to be
\bea
Z_{m_1,0,l_1,l_2}(\scq) & = & \sum_{k=0}^{\min(l_1,l_2)}
\left[ {m_1 + l_1 + l_2 \choose k} - {m_1 + l_1 + l_2 \choose k-1} \right] \nonumber \\
{} & & %\times
{m_1 + l_1 - k \choose m_1}_{\scq^{-1}}
{m_1 + l_2 - k \choose m_1}_{\scq} \;.
\label{z3}
\eea
Again, when evaluated at   $\scq=1$, the  multinomial counting for the  number of $(m_1,0,l_1,l_2)$ biased walks is recovered:
\be
Z_{m_1,0,l_1,l_2}(1) = {m_1+l_1+l_2 \choose m_1,l_1,l_2} \;.
\ee
When also $l_2=0$,  $Z_{m_1,0,l_1,0}(\scq)$ yields, as it should,
the  $\scq$-binomial generating function
\bea
Z_{m, 0, l, 0}(\scq) = {m + l \choose m}_\scq \;
\eea
for the probability distribution of the algebraic area
under staircase walks, ones  that can go only $m$ steps right and $l$ steps up.
One notes that  this $\scq$-binomial has been already introduced in Eq.~(\ref{start})
and evaluated at $\scq$ a root of unity in Eq.~(\ref{identity})
in the context of the cyclic sieving phenomenon
for the $l$-subsets of the $(m+l)$-set.

These  considerations evoke a natural question: Can $Z_{m_1,0,l_1,l_2}(\scq)$ in Eq.~(\ref{z3}) with
$\scq$ a root of unity be an integer, and if so, what does this integer count?
More generally, for $(m_1,m_2,l_1,l_2)$ walks, despite $Z_{m_1,m_2,l_1,l_2}(\scq)$ being generally unknown,
can it be evaluated at $\scq$ a root of unity;
if so, can it yield, at least in certain cases, an integer;
if yes, what does this integer count?

Apart from these cyclic sieving combinatorics/counting considerations, $Z_{m_1,m_2,l_1,l_2}(\scq)$ happens to be 
of interest not only for random walks but also for the quantum Hofstadter problem \cite{Hofstadter},
thanks to the formal mapping \cite{bellissard} between the algebraic area generating function for closed random walks of length $n$
\be\label{ouff} Z_n(\scq)=\sum_{m=0}^{n/2}Z_{m, m, {n\over 2}-m, {n\over 2}-m}(\scq) \ee
and the $n$-moments of the Hofstadter Hamiltonian $H_{\gamma}$
\be  Z_{n}(e^{i\gamma})={\rm Tr}\:H_{\gamma}^{n}\label{ouf} \;,\ee
where 
$\gamma=2\pi\phi/\phi_0$ is the flux per plaquette in units of the flux quantum
and $\scq$ has been taken to be $e^{i\gamma}$. 
Of particular interest is the commensurate case  $\gamma=2\pi p/q$,
with $p$ and $q$ relative primes.
Therefore, another motivation in evaluating $Z_{ m_1, m_2,l_1,l_2}(\scq)$ when $\scq$ is a root of unity
stems from the Hofstadter quantum spectrum itself for a commensurate flux.

In this regard, for ``closed'' biased walks of length $n$ ---
``closed'' here being defined by $l_1=l_2=(n-m_1)/2$
and the choice of particular boundary conditions on the horizontal axis --- an explicit quantum mapping 
has indeed been found  \cite{mo}:
\be \sum_{m=0}^{[\frac{n}{q}]} Z_{qm,0,{n-qm\over 2},{n-qm\over2}}(e^{i\gamma})=
{\rm Tr}\:\tilde{H}_{\gamma}^{n}\label{oufbis},\quad \gamma=2\pi/q \;, \ee
where $\tilde{H}_{\gamma}$ now stands for a truncated Hofstadter Hamiltonian with the horizontal hopping to the left absent. In the sum above, each $Z_{qm,0,(n-qm)/2,(n-qm)/2}$ has to be understood with $n$ and $qm$
being of the same parity  (otherwise it has to be taken equal to 0).
Note also that the number of right steps $qm$ 
is a multiple of $q$, so that one can view the ``closed'' biased walks as  winding on a cylinder with circumference $q$.

\section{The sieving phenomenon and staircase random walks}
As a warm-up, let us rephrase the staircase random walk algebraic area generating function
$Z_{m, 0, l, 0}(\scq)$ evaluated at $\scq$ a root of unity in the context of the
cyclic sieving phenomenon.

One first notes that an $(m, 0, l, 0)$ staircase walk consisting of $m$ steps to the right and $l$ steps up
uniquely corresponds to an $l$-subset of  the set
$\{1, \ldots, m+l\}$. For example, $xxyxxy$, which is one of the possible
$(4, 0, 2, 0)$ walks, corresponds to the subset $\{3,6\}$ of the set $\{1,\ldots,6\}$.
A cyclic permutation $c$ of the above-mentioned set is
$1 \to 2 \to 3 \to \ldots \to m+l \to 1$.
The cyclic sieving phenomenon is the statement that
$Z_{m, 0, l, 0}(\rme^{\frac{2 \rmi \pi p}{m+l}})$
equals the number of the  subsets fixed by $c^{p}$ for $p=1,2,\ldots, m+l$.

Next, one notes that when both $m$ and $l$ are multiples of $q$,
Eq.~(\ref{identity}) implies
\be
Z_{m, 0, l, 0}(\rme^{\frac{2 \rmi \pi}{q}}) = Z_{\frac{m}{q}, 0, \frac{l}{q}, 0}(1) \;,
\label{bb}
\ee
and the quantity on the RHS counts the number of staircase walks with
$m/q$ steps to the right and $l/q$ steps up. 

This counting allows for a simple interpretation in terms of the staircase walk $(m, 0, l, 0)$.
The most general $l$-element subset of the set $\{1, \ldots, m+l\}$ fixed by $c^p$ has the form 
$\{i_1,i_2,\ldots,i_{l/q},i_1+p,i_2+p,\ldots,i_{l/q}+p,\ldots,i_1+(q-1)p,i_2+(q-1)p,\ldots,i_{l/q}+(q-1)p\}$,
where $q = (m+l)/p$ (i.e., $q$ copies of a subset $\{i_1,i_2,\ldots,i_{l/q}\}$, with $i_{l/q} \le p$,
shifted by $p$ with respect to each other).
A subset $\{i_1,i_2,\ldots,i_{l/q}\}$ corresponds, as formulated above, to a
$(\frac{m}{q}, 0, \frac{l}{q}, 0)$ walk.
Thus, any $(m, 0, l, 0)$ walk corresponding to a subset fixed by
$c^p$ is a repetition of $q$ identical ``building blocks'', shifted with respect to each other
by $p$ steps.
Equation (\ref{bb}) expresses the fact that the total number of such walks
is equal to the number of possible building blocks.

Consider an example: $m=9$, $l=3$.
The values of $Z_{9,0,3,0}(\rme^{\frac{2 \rmi \pi  p}{12}})$
with $p$ ranging from 1 to 12 are: \{0, 0, 0, 4, 0, 0, 0, 4, 0, 0, 0, 220\}.
The first 4 in this sequence, corresponding to $p=4$,
counts the number of ways that a $(9,0,3,0)$ walk can be constructed from
3 building blocks, each block being a $(3,0,1,0)$ walk: there are 4 such blocks possible.
Each of those walks is fixed by $c^4$.
The second 4, corresponding to $p=8$, reflects the fact that any subset fixed by $c^4$ is also, trivially,
fixed by $c^8$. 
Finally, every walk is fixed by $c^{12} = I$, so the 220 is the total number of walks
${12 \choose 3}$.
The zero values of $Z_{m,0,l,0}(\rme^{\frac{2 \rmi \pi  p}{m+l}})$ reflect the fact that
there are no walks fixed by $c$, $c^2$, $c^3$, $c^5$, etc.

Note also that the non-zero values can be directly retrieved
from $Z_{m,0,l,0}(\rme^{\frac{2 \rmi \pi p }{q}})$ with $p=1,2,\ldots,q$ where $q$ is the GCD of $m$ and $l$.
In the example above
$Z_{9,0,3,0}(\rme^{\frac{2 \rmi \pi  p}{3}})$
with $p=1,2,3$   directly  yields  \{4, 4, 220\}.

\section{The Stembridge phenomenon for biased and unbiased  random walks}
Consider now in Eq.~(\ref{z3})  the generating function $Z_{m_1,0,l_1,l_2}(\scq)$ 
for the algebraic area distribution of biased walks and evaluate it with $\scq$ a root of unity.
Looking either  at the staircase walks counting in Eq.~(\ref{identity})   with $m$  a multiple of $q$ or  at the $q$-periodic sum in Eq.~(\ref{oufbis}), let $m_1$ be a multiple of $q$.
It follows immediately that $Z_{m_1,0,l_1,l_2}(\rme^{\frac{2 \rmi \pi }{q}})$ is an integer.
Furthermore, as for staircase walks\footnote{The analogy stops here: 
$Z_{m_1,0,l_1,l_2}(\rme^{2 \rmi \pi p\over m_1+l_1+l_2})$  with $p=1,2,\ldots,m_1+l_1+l_2$
would yield non-integer values for certain values of $p$.},
all $Z_{m_1,0,l_1,l_2}(\rme^{\frac{2 \rmi \pi p}{q}})$'s  for  $p=1,2,\ldots,q$
are also non-vanishing integers.

One can go a step further by having not only $m_1$ but also $|l_1-l_2|$
be a multiple of $q$, as suggested by Eq.~(\ref{bb}). Some algebra allows
then to rewrite $Z_{m_1,0,l_1,l_2}(\rme^{\frac{2 \rmi \pi }{q}})$ as
\bea
Z_{m_1,0,l_1,l_2}(\rme^{\frac{2 \rmi \pi }{q}}) = && \sum_{k=\min(l_1,l_2)}^{0,-q}  {m_1 + l_1 + l_2 \choose k}\frac{m_1(m_1 + l_1+ l_2-2k)}{(m_1 + l_1-k)(m_1 + l_2-k)} \nonumber \\
&& {} {\frac{m_1 + l_1-k}{q} \choose \frac{m_1}{q}}
{\frac{m_1 + l_2-k}{q} \choose \frac{m_1}{q}}
 \;. \label{z3q}
\eea
In Eq.~(\ref{z3q}), $\sum_{k=\min(l_1,l_2)}^{0,-q}$    means that the sum over $k$ is from $\min(l_1,l_2)$  down to 0 by steps of  minus $q$ --- from which one deduces that the entries of the  last two binomials are indeed integers.
It is also understood that when $m_1=0$, the ratio
${m_1(m_1 + l_1+ l_2-2k)\over (m_1 + l_1-k)(m_1 + l_2-k)}$ is non-vanishing
and actually equal to $1$ only when $k=\min(l_1,l_2)$.  This yields, as it should,
$Z_{0,0,l_1,l_2}(\rme^{\frac{2 \rmi \pi }{q}}) =  { l_1 + l_2 \choose\min(l_1,l_2) }={ l_1 + l_2 \choose l_1 }$, the number of  $(l_1,l_2)$ walks on the vertical axis which all have, trivially,
a vanishing algebraic area so that $Z_{0,0,l_1,l_2}(\rme^{\frac{2 \rmi \pi }{q}})=Z_{0,0,l_1,l_2}(1)$.

In (\ref{z3q}) for each\footnote{Note that for each such $k$,
\[\frac{m_1(m_1 + l_1+ l_2-2k)}{(m_1 + l_1-k)(m_1 + l_2-k)}{\frac{m_1 + l_1-k}{q} \choose \frac{m_1}{q}}
{\frac{m_1 + l_2-k}{q} \choose \frac{m_1}{q}}\]
is by itself an integer, which is nothing but saying that, trivially,   
\[\frac{a(a + b+ c)}{(a + b)(a + c)}{a + b \choose a}
{a + c \choose a}\]
is an integer for any integers $a,b,c$.}  $k$    the first binomial counts the number of ways to pick $k$ elements out of the 
$(m_1 + l_1 + l_2)$-element set, whereas the last two binomials
 are the  numbers of $(\frac{m_1}{q}, \frac{l_1-k}{q})$
and $(\frac{m_1}{q}, \frac{l_2-k}{q})$ staircase walks, respectively. 
Finally, the ratio in (\ref{z3q}) encodes the fact that the $l_1-k$
steps up and $l_2-k$ steps down ``annilihate'' each other
on the vertical axis so that these staircase walks cannot be considered as independent. 

Now, let us focus  on the Stembridge's case $\scq = -1$,  that is to say $q=2$:
one can then simplify (\ref{z3q}) to get the quite  convincing expression
\be
Z_{m_1,0,l_1,l_2}(-1)=\binom{{l_1}+{l_2}}{l_1} \binom{\frac
   {m_1+{l_1}+{l_2}}{2}}{\frac{{l_1}+{l_2}}{2}} \;.
\label{z3-1}
\ee
Furthermore, one can go a step further and relax the biased constraint $m_2=0$, that is to say
consider now more general $(m_1,m_2,l_1,l_2)$ walks,
with  both $|m_1-m_2|$ and $|l_1-l_2|$  multiples of $q$.
Again, in the Stembridge's case  $q=2$,  one obtains 
\be\label{good}
Z_{m_1,m_2,l_1,l_2}(-1) = {m_1 + m_2 \choose m_1}{l_1 + l_2 \choose l_1}{\frac{m_1+m_2+l_1 + l_2}{2} \choose \frac{l_1+l_2}{2}} \;,
\ee
which does reduce to Eq.~(\ref{z3-1}) for $m_2=0$ biased walks.

Eq.~(\ref{good}),  and consequently Eq.~(\ref{z3-1}),   yields integers, since  $|m_1-m_2|$ and $|l_1-l_2|$ being even implies that
$m_1+m_2$ and $l_1+l_2$ are even as well.
A  Stembridge combinatorial interpretation of these integers  is depicted in Fig.~1.
Divide the $(m_1 + m_2 + l_1 + l_2)$ cells into two equal subsets
($q = 2$, i.e., two building blocks).
Then consider four types of objects (corresponding to four directions --- up, down, left, right)
and define a subset as ``fixed in the weak sense'' by $c^{(m_1 + m_2 + l_1 + l_2)/2}$ --- meaning that
when acting by $c$, ``up'' is considered to be the same as ``down'' and
``left'' the same as ``right''.
The formula is then interpreted as follows:
the rightmost factor counts the number of ways that half of the (up + down) objects,
$(l_1 + l_2)/2$,
can be distributed between half the cells, $(m_1 + m_2 + l_1 + l_2)/2$.
Then two other factors distinguish between ``up'' and ``down'' ($l_1$ out of $l_1 + l_2$) and
between ``left'' and ``right'' ($m_1$ out of $m_1 + m_2$).
\begin{figure}
\begin{center}
\includegraphics[scale=.3,angle=0]{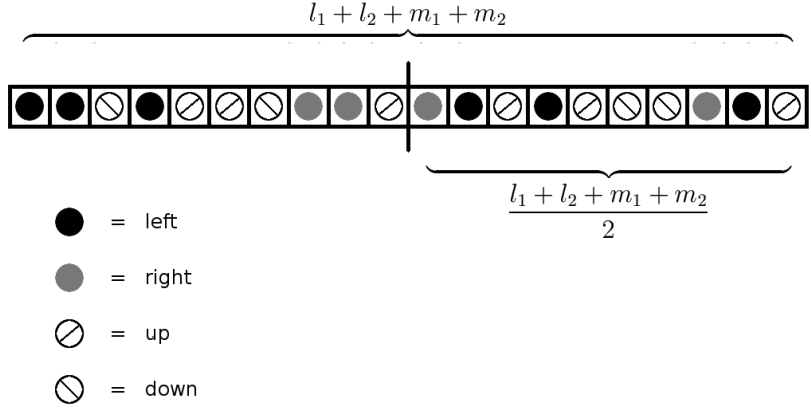}\label{fig2}
\caption{A Stembridge interpretation.}
\end{center}
\end{figure}

The simple expression of $ Z_{m_1,m_2,l_1,l_2}(\rme^{\frac{2 \rmi \pi }{2}})$ in Eq.~(\ref{good}) asks for a generalization to $ Z_{m_1,m_2,l_1,l_2}(\rme^{\frac{2 \rmi \pi }{q}})$ for  $(m_1,m_2,l_1,l_2)$ walks with  both $|m_1-m_2|$ and $|l_1-l_2|$  multiples of $q$.
Such an expression for $ Z_{m_1,m_2,l_1,l_2}(\rme^{\frac{2 \rmi \pi }{q}})$   would  in turn allow one to retrieve the $n$-moments of the Hofstadter  Hamiltonian at commensurate flux by summing $Z_{m,m,l=n/2-m,l=n/2-m}(\scq)$
over $m$, like in Eq.~(\ref{ouff}) --- with in that case  both $|m_1-m_2|$ and $|l_1-l_2|$  vanishing,
in principle an even more simple situation --- and then using Eq.~(\ref{ouf}).
Following this line of reasoning for $q=2$, i.e., the $\scq=-1$ Stembridge case,  one eventually gets from Eq.~(\ref{good}) the $n$th moment of $H_{\pi}$ 

\be\label{nicenice} \tr H_{\pi}^n=\sum_{m=0}^{n/2} Z_{m,m,l=n/2-m,l=n/2-m}(-1)=\sum_{m=0}^{n/2}{2 m\choose m} {2 (\frac{n}{2} - m)\choose \frac{n}{2} - m}{\frac{n}{2}\choose m} \;. \ee

This last result is a sum rule for the Hofstadter spectrum at commensurate flux $\gamma=\pi$. Below we  compute $\tr H_{2\pi/q}^n$
  for  particular values of $q=2,3,4,6$ directly from the quantum mechanics formulation (i.e., without relying on an evaluation of $ Z_{m,m,n/2-m,n/2-m}(\rme^{\frac{2 \rmi \pi }{q}})$).

\section{The $n$th moments of the Hofstadter Hamiltonian}

Up to now we have used the mapping of the quantum Hofstadter problem on classical random walks to arrive at the Hofstadter sum rule (\ref{nicenice}). This sum rule can be directly retrieved  from   the quantum Hofstadter Hamiltonian itself, as can be
other sum rules for simple values of  $q=3, 4, 6$. 

For concreteness, we wish to compute the $n$th moment
\be\label{trace}
\tr H_\gamma^n%=Z_n(e^{i\gamma})
\ee
where $H_\gamma$ is the Hofstadter Hamiltonian
\be
H_\gamma = x+x^{-1} +y +y^{-1} ~,~~~ xy  = e^{i\gamma} yx \;;
\ee
$x$ and $y$ stand now for the quantum lattice hopping operators on the horizontal and vertical directions respectively:
they do non commute, due to the presence of the perpendicular magnetic field with flux $\gamma$ per unit cell.
The definition of trace in (\ref{trace}) is such that
\be
\tr ( x^m y^n ) = \delta_{m,0} \, \delta_{n,0} \;.
\ee

We define the generating function
\be
{\cal Z} (\rme^{\rmi\gamma},t) = \tr\sum_{n=0}^\infty H_{\gamma}^n  \, t^n = \tr \frac{1}{1-tH_\gamma} = \tr \frac{1}{1-t^2 H_\gamma^2}
\ee
where the last rewriting follows from the vanishing of $\tr H_\gamma^n$ for odd $n$.

From now on we focus on the commensurate case  $\gamma=2\pi/q$,
for which  the operators $x^q$ and $y^q$ do commute with $x$ and $y$ and are Casimirs,
and so is
\be
C_q = x^q + x^{-q} + y^q + y^{-q} \;.
\ee
$C_q$ is, essentially, the Hofstadter Hamiltonian with $q=1$, i.e., $\gamma=2\pi$, that is to say  zero flux.
One easily obtains
\be
\tr C_q^{2n} = {{2n}\choose{n} }^2 
\label{Cex}
\ee
and
\bea
\tr H_{2\pi/q}^2  &=& 4 \;, \cr
\tr H_{2\pi/q}^4  &=& 28 + 8\cos(2\pi/q) \;, \cr
\tr C_q^{2n+1}  &=& \tr H_{2\pi/q}^{2n+1} = 0 \;,
\eea
so $\tr H_{2\pi/q}^{n}$ will be nonzero only for even values of $n$, as expected.

The operators $x$, $y$ can be represented as $q \times q$ matrices (with $x^q$ and $y^q$ 
proportional to the identity matrix) and therefore so can $H_{2\pi/q}$. As a result, it satisfies a characteristic
equation of degree $q$, and so $H_{2\pi/q}^q$ can be expressed in terms of lower powers of $H_{2\pi/q}$
and the Casimir $C_q$. Defining the ``parity" of the monomial $x^m y^n$ as $(-1)^{m+n}$, $H_{2\pi/q}^n$
has parity $(-1)^n$ and only like-parity powers of $H_{2\pi/q}$ will appear in the expression of $H_{2\pi/q}^q$.
Overall we have
\be
H_{2\pi/q}^q = C_q+ 2q H_{2\pi/q}^{q-2} + c_{q-4} H_{2\pi/q}^{q-4} +  \cdots + c_{q({\rm mod}~2)} H_{2\pi/q}^{q({\rm mod}~2)} \;,
\ee
where $c_k$ are $q$-dependent numerical coefficients.
Looking back at random walks, this is a rewriting of Eq.~(\ref{qbingen}) but with the different terms regrouped into powers of
$H_{2\pi/q}$ itself, which is possible for a rational flux $\gamma={{2\pi}/{q}}$.
The term $C_q$ arises out of the terms $x^q$, $x^{-q}$, $y^q$, $y^{-q}$ in the expansion of
$H_{2\pi/q}^q$. The fact that the coefficient $c_{q-2}$ of the term $H_{2\pi/q}^{q-2}$ is $2q$ can be proven using
combinatorics and the identity $1+e^{2\rmi\pi/q} + \cdots +(e^{2\rmi\pi/q})^{q-1} = 0$.

Performing the above rewriting explicitly for the first few values of $q$, we obtain
\bea
q=2:&~~~H_\pi^2 &= 4+C_2 \cr
q=3:&~~~H_{2\pi/3}^3 &= 6H_{2\pi/3} + C_3 \cr
q=4:&~~~H_{\pi/2}^4 &= 8 H_{\pi/2}^2 + C_4 -4 \cr
q=5:&~~~H_{2\pi/5}^5 &= 10 H_{2\pi/5}^3 -(5-2\cos{2\pi\over  5}) H_{2\pi/5} + C_5 \cr
q=6:&~~~H_{\pi/3}^6 &= 12 H_{\pi/3}^4 -24 H_{\pi/3}^2 +4+C_6
\eea
The above relations will allow for the exact evaluation of $\tr H_{2\pi/q}^{n}$. The zero flux case $q=1$, i.e., $\gamma=2\pi$ (corresponding to the unweighted random
walk, $\scq=1$) is explicitly solved as
\be
\tr H_{2\pi}^{2n}= {{2n}\choose{n} }^2 ,~~~ {\cal Z} (1,t) = \frac{2}{\pi} K(4t)
\ee
with $K(k)$ the complete elliptic integral of the first kind.

The simplest nontrivial flux is $q=2$, i.e., $\gamma=\pi$, that is, in the language of random walks, the Stembridge case $\scq=-1$. It
serves to demonstrate a common phenomenon, the appearance of various different-looking
expressions for the same quantity related through combinatorial identities. For the generating function we have
\bea
{\cal Z}(-1,t) &=& \tr \frac{1}{1-t^2 H_\pi^2} = \tr \frac{1}{1-t^2 (4+C_2 )} \cr
&=&
\frac{1}{1-4t^2} \tr \frac{1}{1-\frac{t^2}{1-4t^2} C_2} = \frac{2}{\pi} \frac{1}{1-4t^2}
K\left(\frac{4t^2}{1-4t^2}\right) \;,
\eea
where we have used the fact that $C_q$ is essentially the free Hamiltonian $H_{2\pi}$.
Equivalently,
\bea
\tr H_\pi^{2n}  &=&   \tr (4+C_2 )^n = \sum_{k=0}^{[n/2]} {n \choose 2k} 4^{n-2k} \tr C_2^{2k}\cr
&=& \sum_{k=0}^{[n/2]} 2^{2n-4k} {n\choose 2k}{2k \choose k}^2 \;.
\eea

An alternative formula is obtained by writing $H_\pi^2$ as a sum of two commuting parts,
\be
H_\pi^2 = 4+C_2 = 4 + x^2 + x^{-2} + y^2 + y^{-2} = (x + x^{-1})^2 + (y + y^{-1})^2
\ee
and thus
\bea
\tr H_\pi^{2n}  &=& \tr \left[ (x + x^{-1})^2 + (y + y^{-1})^2 \right]^n \cr
&=& \sum_{k=0}^n {n \choose k} \tr (x + x^{-1})^{2k}  \, \tr (y + y^{-1})^{2(n-k)} \cr
&=&
 \sum_{k=0}^n {n\choose k}{2k\choose k}{2n-2k\choose n-k} \;,
\eea
recovering Eq.~(\ref{nicenice}).

Yet another formula is obtained by noticing that the value of the Casimir $C_2$ is between
$-4$ and $4$ (since $x^2$ and $y^2$ are phases) and thus it makes sense to put
\be
\cos \phi = \frac{C_2}{4}
\label{cos}
\ee
with some angle $\phi$, it terms of which
\be
H_\pi^2 = 4 (1+\cos \phi ) = 8 \cos^2 \frac{\phi}{2} = 2 \left( e^{i\frac{\phi}{2}} + e^{-i\frac{\phi}{2}} \right)^2 \;.
\ee
In this representation
\be
H_\pi^{2n} = 2^n \left( e^{i\frac{\phi}{2}} + e^{-i\frac{\phi}{2}} \right)^{2n}
= 2^n \sum_m {2n \choose m} e^{i(n-m)\phi} = 2^n \sum_m {2n \choose m} \cos (n-m)\phi \;,
\ee
where in the last step we used the fact that only the real part of $e^{i(n-m)\phi}$ contributes.
Changing the variable $m = n+k$, $-n \le k \le n$, using the formula
\be
\cos k \phi = \sum_{l =0}^{\left[\frac{k}{2}\right]} 2^{k-2l-1} (-1)^l \frac{k}{k-l}
{k-l \choose l} (\cos\phi )^{k-2l} ~,~~~k\ge 1
\ee
and using Eq.~(\ref{cos}), we obtain
\be
H_\pi^{2n} = \sum_{k=1}^n \sum_{l=0}^{[k/2]} 2^{n-k+2l} (-1)^l \frac{k}{k-l} 
{2n \choose n+k} {k-l \choose l} C_2^{k-2l}
\ee
(in the above, $\frac{k}{k-l}$ is defined to be $1$
when both numerator and denominator are zero, as well as the ratios $\frac{2k}{2k-l}$ or  $\frac{l}{l-s}$ below.)
Taking the trace and using Eq.~(\ref{Cex}), only even $k$ values will contribute, and we obain a third
formula for $q=2$:
\be \tr H_\pi^{2n} =
\sum_{k=0}^{[n/2]}  \sum_{l=0}^k (-1)^l \, 2^{n-2k+2l} {2n \choose n+2k}\frac{2k}{2k-l}
 {2k-l \choose l} {2k-2l \choose k-l}^2 \;.
\label{Z23}
\ee

It is remarkable that the above three different-looking formulas all agree and give the same result.
The last formula may look a bit unmotivated and of dubious value, as it is a double sum (rather than
a simple sum like the first two), but it is given in order to make contact with the upcoming results for higher
$q$.

For the case $q=3$, the derivation proceeds by solving the characteristic equation for $H_{{2\pi}/{3}}$,
$\lambda^3 = 6\lambda +C_3$. The solutions are
\be
\lambda_p = 2\sqrt{2}  \cos \frac{\phi + 2p\pi}{3}  ~,~~~ p=0,1,2
\ee
with the angle $\phi$ now satisfying
\be
\cos \phi = \frac{C_3}{4\sqrt 2} \;.
\ee
Evaluating $\tr H_{{2\pi}/{3}}^n$ amounts to averaging $\lambda_p^n$ over the three values of $p$.
In a calculation analogous to the third formula for $q=2$ we have
\bea
\tr H_{2\pi/3}^{2n} & = & \frac{1 }{3}\tr \sum_{p=0,1,2} \lambda_p^{2n} =  \frac{2^n}{3}  \sum_{p=0,1,2} \tr \left( \rme^{\rmi\frac{\phi + 2p\pi}{3}}
+ \rme^{-\rmi\frac{\phi + 2p\pi}{3}} \right)^{2n} \nonumber \\
& = & \frac{2^n}{3}  \sum_{p,m} 
{2n \choose m} \tr \rme^{\rmi 2(n-m)\frac{\phi + 2p\pi}{3}} \;.
\eea
The sum over $p$ above will give zero unless $n-m$ is a multiple of 3. Putting $n-m=3k$, we have
\be
\tr H_{2\pi/3}^{2n}= 2^n \sum_k
{2n \choose n+3k}\tr \rme^{\rmi2k\phi} =  2^n \sum_k {2n \choose n+3k} \tr \cos (2k\phi ) \;.
\ee
The rest of the calculation proceeds in a way similar to the derivation of the last formula
for $q=2$. The result is
\be
\tr H_{{2\pi}/{3}}^{2n}= \sum_{k=0}^{[n/3]} \sum_{l=0}^k 
 (-1)^l \, 2^{n-3k+3l} {2n \choose n+3k}\frac{2k}{2k-l}
{2k-l \choose l} {2k-2l \choose k-l}^2 \;.
\label{Z3}
\ee
The similarity with (\ref{Z23}) for $q=2$ is obvious.

For $q=4$ we proceed in a similar way. The solutions to the characteristic equation $\lambda^4 =
8 \lambda^2 + 4 -C_4$ are
\bea
\lambda_p &=& \pm \sqrt{4 \pm \sqrt{12+C_4}} \cr
&=& 2 \sqrt{2} \cos \frac{\phi + 2p\pi}{4} ~,~~ p=0,1,2,3
\eea
with
\be
\cos \phi = \frac{4+C_4}{8} \;.
\ee
Expressing $\tr H_{\pi/2}^{2n}= \frac{1}{4} \sum_p \lambda_p^{2n}$ and choosing either the first or the
second expression above for $\lambda_p$ leads to two different-looking formulas. The first is
\be
\tr H_{{\pi}/{2}}^{2n} = \sum_{k=0}^{[n/2]} \sum_{l=0}^{[k/2]} 2^{2n-2k-4l} \, 3^{k-2l} \, 
{n \choose 2k}{k\choose 2l}{2l \choose l}^2 \;.
\ee
The second one proceeds, again, in a way similar to the last of $q=2$ and the $q=3$ cases
and gives
\be
\tr H_{{\pi}/{2}}^{2n} =  %2^n {2n \choose n} +
\sum_{k=0}^{[n/2]} \sum_{l=0}^{[k/2]} \sum_{s=0}^{[k/2]-l}  
(-1)^l \, 2^{n-4s} {2n\choose n+2k} \frac{k}{k-l}
{k-l \choose l}{k-2l\choose 2s} {2s \choose s}^2 \;.
\ee
This is similar to (\ref{Z23}) and (\ref{Z3}) but with a different structure. Clearly, the fact that $q=4$ is not prime
is related to the different appearance of the result.

The case $q=5$ presents some qualitatively new features. It is the first case for which the
characteristic equation involves non-rational coefficients. Further, since this equation is
quintic, it has (in principle) no analytical solutions. We will present no explicit formula for
$\tr H_{2\pi/5}^{2n}$  and leave its full treatment for a future publication.

Finally, we deal with the simpler $q=6$ case. The solutions to the characteristic equation are now
\be
\lambda_p = \pm 2 \sqrt{1+ \sqrt{2} \cos\frac{\phi+2p\pi}{3}} ~,~~ p=0,1,2
\ee
with
\be
\cos \phi = \frac{36+C_6}{32 \sqrt 2} \;.
\ee
A calculation along the lines of the previous ones gives
\bea
\tr H_{{\pi}/{3}}^{2n}&= &\sum_{k=0}^{[n/2]} \sum_{l=1}^{\left[\frac{n-2k}{3}\right]} \sum_{s=0}^{[l/2]} \sum_{t=0}^{[l/2]-s}
(-1)^s ~ 2^{2n - k -4l +5s -4t} ~ 3^{2l-4s-4t}\cr
&&{n\choose 2k+3l}{2k+3l \choose k} \frac{l}{l-s} {l-s \choose s} {l-2s \choose 2t}
{2t \choose t}^2  \;.
\eea
which is  even more elaborate than the previous formulas.

This kind of computation can in principle be extended to higher values of $q$. Overall, we see no
discernible pattern in the formulas obtained so far for $\tr H_{2\pi/q}^{2n}$.
However, the fact that multiple not manifestly equivalent expressions
exist for each $q$ leaves the hope that a form more amenable to generalization may still exist.

\hspace{1cm}

\section{Conclusion}

To conclude, we have further explored the previously-considered
generating function $Z_{m_1,m_2,l_1,l_2}(\scq)$ of the probability distribution
of the algebraic area of random walks on a square lattice.
We have: (i) elucidated the relation of $Z_{m,0,l,0}(\rme^{2\rmi\pi/q})$ (``staircase walks'')
to the cyclic sieving phenomenon (by demonstrating that its values
correspond to the numbers of subsets fixed by powers of the cyclic permutation);
(ii) shown that the values of $Z_{m_1,0,l_1,l_2}(\rme^{2\rmi\pi/q})$ (``biased walks'') are integers under
certain conditions, and explained why;
(iii) obtained an explicit expression for $Z_{m_1,m_2,l_1,l_2}(-1)$ (Stembridge case),
and connected it with the expression for the $n$-th moment of the Hofstadter Hamiltonian
at the simplest nontrivial flux, $q=2$;
(iv) found explicit expressions for the same $n$-th moment for $q=3, 4, 6$.
Obtaining either a general closed expression for  $Z_{m_1,m_2,l_1,l_2}(\rme^{2\rmi\pi/q})$ or at least more
particular cases thereof, as well as a general form of the $n$-th moment
evaluated at any root of unity, are open tasks. Note that the values of
$Z_{m_1,m_2,l_1,l_2}(\rme^{2\rmi\pi/q})$ are seen numerically to be not integer
anymore when $q\ge 5$ (to the exception of $q=6$), making the task even more challenging.
\vskip 0.4cm
\noindent
{\bf Acknowledgements:}
\vskip 0.2cm
S.O. would like to thank F. Breuer and S. Wagner  for drawing his attention to Ref.~\cite{sieving}  and for early discussions on the subject. He would also like to thank the City College of New York for the warm hospitality in the spring of 2014 when part of this work was done. A.P.'s research is supported by NSF grant 1213380 and by a PSC-CUNY grant.

\end{document}